\documentclass[aps,prd,preprint,nofootinbib]{revtex4}
\usepackage{amsfonts}
\usepackage{mathrsfs}
\usepackage{graphicx}
\usepackage{amsmath}
\usepackage{amssymb}
\usepackage{subfigure}
\usepackage{epsfig}
\usepackage{graphicx}
\usepackage{slashed}
\usepackage{color}
\usepackage{ulem}
\newcommand{\bqa}{\begin{eqnarray}}
\newcommand{\eqa}{\end{eqnarray}}
\newcommand{\beq}{\begin{equation}}
\newcommand{\eeq}{\end{equation}}
\allowdisplaybreaks[1]
\graphicspath{{fig/}{dia/}} \DeclareGraphicsExtensions{.eps}
\begin{document}

\title{Finding $B_c(3S)$ States via Their Strong Decays\\[0.7cm]}

\author{\vspace{1cm} Rui Ding$^1$, Bing-Dong Wan$^1$, Zi-Qiang Chen$^{1,3}$, Guo-Li Wang$^2$\footnote[1]{gl\_wang@hit.edu.cn}, \\
 Cong-Feng Qiao$^{1,3}$\footnote[2]{qiaocf@ucas.ac.cn} \\}

\affiliation{$^1$ School of Physics, University of Chinese Academy of
Sciences, Beijing 100049, China\\
$^2$ Department of Physics, Hebei University, Baoding 071002, China\\
$^3$ CAS Center for Excellence in Particle Physics, Beijing 10049, China
\vspace{0.6cm}}

\begin{abstract}
\vspace{0.5cm}

The experimentally known $B_c$ states are all below open bottom-charm threshold, which experience three main decay modes, and all induced by weak interaction. In this work, we investigate the mass spectrum and strong decays of the $B_c(3S)$ states, which just above the threshold, in the Bethe-Salpeter formalism and $^3P_0$ model.
The numerical estimation gives $M(B_c(3^1S_0))=7273\ {\rm MeV}$, $M(B_c^*(3^3S_1))=7304\ {\rm MeV}$, $\Gamma\left(B_c(3^1S_0)\to B^*D\right)=26.02^{+2.33}_{-2.21}\ {\rm MeV}$, $\Gamma\left(B_c^*(3^3S_1)\to BD\right)=3.39^{+0.27}_{-0.26}\ {\rm MeV}$, $\Gamma\left(B_c^*(3^3S_1)\to B^*D\right)=14.77^{+1.40}_{-1.33}\ {\rm MeV}$ and $\Gamma\left(B_c^*(3^3S_1)\to BD^*\right)=6.14^{+0.58}_{-0.54}\ {\rm MeV}$.
Compared with previous studies in non-relativistic approximation, our results indicate that the relativistic effects are notable in
$B_c(3S)$ exclusive strong decays. According to the results, we suggest to find the $B_c(3S)$ states in their hadronic decays to $B$ and $D$ mesons in experiment, like the LHCb.

\end{abstract}
 \maketitle
\newpage

The $B_c$ meson family is unique in quark model as its states are composed of heavy quarks with different flavors.
The $B_c$ mesons lie intermediate between ($c\bar{c}$) and ($b\bar{b}$) states both in mass and size, while the different quark masses leads to much richer dynamics.
On the other hand, the $B_c$ mesons cannot annihilate into gluons or photons and thus they are very stable.
The $B_c$ mesons provide a unique window to reveal information about heavy-quark dynamics and can deepen our understanding of both   strong and weak interactions.

Although there have been many investigations in the literature \cite{Eichten:1980mw,Eichten:1994gt,Eichten:2019gig,Godfrey:1985xj,Godfrey:2004ya,Kiselev:1994rc,Zeng:1994vj,Kiselev:1996un,Ebert:2002pp,AbdElHady:2005bv,Ferretti:2015rsa,Mathur:2018epb,Li:2019tbn,Asghar:2019qjl,Ortega:2020uvc,Guleria:2020kuy,Abreu:2020ttf,Narison:2020wql,Berezhnoy:2021ipt,Aliev:2019wcm,Chang:2019eob} about the properties of $B_c$ mesons, the excited $B_c$ states, especially above threshold, are rarely explored.
The ground state $B_c$ meson was first observed by the CDF Collaboration at Fermilab~\cite{Abe:1998fb} in 1998, while there was no reported evidence of the excited $B_c$ state until 2014, the ATLAS Collaboration reported a structure with mass of $6842\pm 9$ MeV~\cite{Aad:2014laa}, which is consistent with the value predicted for $B_c(2S)$.
Recently, the excited $B_c(2^1S_0)$ and $B_c^\ast(2^3S_1)$ states have been observed in the $B_c^+\pi^+\pi^-$ invariant mass spectrum by the CMS and LHCb Collaboration, with their masses determined to be $6872.1\pm 2.2$ MeV and $6841.2\pm1.5$ MeV \cite{Sirunyan:2019osb,Aaij:2019ldo}, respectively.
Since the low-energy photon in the intermediate decay $B_c^\ast\to B_c\gamma$ was not reconstructed, the mass of $B_c^\ast(2^3S_1)$ meson appears lower than that of  $B_c(2^1S_0)$.

The successful observation of $B_c(2S)$ states stimulates the interest in searching for $B_c(3S)$ states.
Motivated by this, in this work, we calculate the mass spectrum of $B_c(nS)$ states up to $n=4$ in the framework of Bethe-Salpeter (BS) equation \cite{s5}.
A long-ranged linear confining potential and a short-ranged one gluon exchange potential are used in our calculation.
Our results indicate that the $B_c(3S)$ states lie above the threshold for decay into a $BD$ meson pair.
By combining $^3P_0$ model with the calculated relativistic BS wave functions, we investigate the strong decay properties of $B_c(3S)$ mesons.
We also estimate the corresponding numbers of events at the Large Hadron Collider (LHC) experimental condition.
Although similar topic has been studied in non-relativistic framework \cite{Eichten:2019gig,Kiselev:1996un,Ferretti:2015rsa,Li:2019tbn,Asghar:2019qjl}, the relativistic treatment of $B_c(3S)$ exclusive decays is evidently more close to the reality.

In quantum field theory,  the BS equation provides a basic description for bound states.
The BS wave function of a quark-antiquark bound state is defined as
\begin{equation}
\label{e1}
\chi(x_1,x_2)=\langle 0| T\psi(x_1)\bar{\psi}(x_2)| P\rangle,
\end{equation}
where $x_1$ and $x_2$ are the coordinates of the quark and antiquark respectively, $P$ is the momentum of the bound state, $T$ denotes the time ordering operator.
 The wave function in momentum space is
\begin{equation}
\label{e2}
\chi_P(q)=e^{{-iP}\cdot X}\int d^{4}x e^{{-iq}\cdot x} \chi(x_1,x_2),
\end{equation}
where $q$ is the relative momentum between the quark and antiquark.
The ``center-of-mass coordinate'' $X$ and the ``relative coordinate'' $x$ are defined as:
\begin{equation}
X = \frac{m_1}{m_1+m_2} x_1 + \frac{m_2}{m_1+m_2} x_2,~~~x=x_1 - x_2,
\end{equation}
where $m_1$ and $m_2$ are the masses of the quark and antiquark respectively.
Then the bound state BS equation in momentum space reads
\begin{equation}
\label{e3}
S^{-1}_1(p_1)\chi_{_P}(q)S^{-1}_2(-p_2)=i\int\frac{d^4k}{(2\pi)^4} V(P; q,k)\chi_{_P}(k).
\end{equation}
Here $S_i(\pm p_i)=\frac{i}{\pm\slashed{p}_i-m_i}$ denotes the fermion propagator;
$V(P; q,k)$ is the interaction kernel;
$p_1$ and $p_2$ are the momenta of the quark and anti-quark respectively, which can be expressed as
\begin{equation}
\label{e4}
p_{i} = \frac{m_{i}}{m_{1}+m_{2}}P+Jq,
\end{equation}
where,  $J=1$ for the quark ($i=1$) and $J=-1$ for the antiquark ($i=2$).
With the definitions $p_{i_P} \equiv \frac{P\cdot p_i}{M}$ and  $p^\mu_{i\perp} \equiv p_i^\mu - \frac{P\cdot p_i}{M^2}P^\mu$, the propagator $S_i(Jp_i)$ can be decomposed as
\begin{equation}
-iJS_i(Jp_i)=\frac{\Lambda^+_i(q_{\perp})}{p_{i_P}-\omega_i+i\epsilon}+\frac{\Lambda_i^-(q_{\perp})}{p_{i_P}+\omega_i-i\epsilon},
\end{equation}
where
\begin{equation}
\label{projector}
\begin{aligned}
&\Lambda_i^{\pm}(q_{\perp})\equiv\frac{1}{2\omega_i}\left[\frac{\slashed P}{M}\omega_i\pm(\slashed p_{i\perp} + J m_i)\right],\\
&\omega_i \equiv \sqrt{m_i^2 - p_{i\perp}^2}.
\end{aligned}
\end{equation}

Under the instantaneous approximation, the interaction kernel in the center of mass frame takes the form $V(P; q, k)|_{\vec{P}=0} \approx V(q_\perp, k_\perp)$. Then the BS equation can be reduced to
\begin{equation}
\label{bswf}
\begin{aligned}
\chi_{_P}(q)=S_1(p_1)\eta_{_P}(q_\perp)S_2(-p_2),
\end{aligned}
\end{equation}
with
\begin{equation}
\label{eta}
\eta_{_P}(q_\perp)=\int\frac{d^3k_\perp}{(2\pi)^3} V(q_\perp,k_\perp)\varphi_{_P}(k_\perp),
\end{equation}
where $\varphi_{_P}(q^\mu_\perp)\equiv i\int\frac{dq_{_P}}{2\pi}\chi_{_P}(q)$ is the 3-dimensional BS wave function.
By introducing the notation $\varphi^{\pm\pm}_{_P}(q_{\perp})$ as:
\begin{equation}
\varphi^{\pm\pm}_{_P}(q_{\perp})\equiv
\Lambda^{\pm}_{1}(q_{\perp})
\frac{\slashed{P}}{M}\varphi_{_P}(q_{\perp})
\frac{\slashed{P}}{M} \Lambda^{{\pm}}_{2}(q_{\perp}),
\label{eq10}
\end{equation}
the wave function can be decomposed as
\begin{equation}
\varphi_{_P}(q_{\perp})=\varphi^{++}_{_P}(q_{\perp})+
\varphi^{+-}_{_P}(q_{\perp})+\varphi^{-+}_{_P}(q_{\perp})
+\varphi^{--}_{_P}(q_{\perp}).
\end{equation}
And the BS equation \eqref{bswf} can be decomposed into four equations
\begin{align}\label{varphi++}
&(M-\omega_{1}-\omega_{2})\varphi^{++}_{_P}(q_{\perp})=
\Lambda^{+}_{1}(q_{\perp})\eta_{_P}(q_{\perp})\Lambda^{+}_{2}(q_{\perp}), \\
&(M+\omega_{_1}+\omega_{2})\varphi^{--}_{p}(q_{\perp})=-
\Lambda^{-}_{1}(q_{\perp})\eta_{_P}(q_{\perp})\Lambda^{-}_{2}(q_{\perp}),\\
&\varphi^{+-}_{_P}(q_{\perp})=\varphi^{-+}_{_P}(q_{\perp})=0.
\label{eq11}
\end{align}

To solve the BS equation, one must have a good command of the potential between two quarks.
According to lattice QCD calculations, the potential for a heavy quark-antiquark pair in the static limit is well described by a long-ranged linear confining potential (Lorentz scalar $V_S$) and a short-ranged one gluon exchange potential (Lorentz vector $V_V$) \cite{la,lb,Qiao:1996re}:
\begin{align}
\label{a9}
&{V(r)}={V_S(r)+\gamma_\mu\otimes\gamma^\mu V_V(r)},\nonumber \\
&{V_S(r)}={\lambda r\frac {(1-e^{-\alpha r})}{\alpha
r}}+V_0,\nonumber \\
&{V_V(r)}=-{\frac 43}{\frac {\alpha_{s}(r)} r}e^{-\alpha r}.
\end{align}
Here, the factor $e^{-\alpha r}$ is introduced not only to avoid the infrared divergence but also to incorporate the color screening effects of the dynamical light quark pairs on the ``quenched'' potential \cite{t13}.
The potentials in momentum space are
\begin{align}
\label{a12}
&V( \vec{p})=(2\pi)^3V_S( \vec{p}) +\gamma_\mu\otimes \gamma^\mu
(2\pi)^3V_V( \vec{p}),\nonumber \\
&V_S( \vec{p})=-\Big(\frac \lambda \alpha+V_0\Big)
\delta ^3( \vec{p})+\frac \lambda {\pi
^2}\frac 1{( \vec{p}^{\;2}+\alpha ^2)
^2},\nonumber \\
&V_V( \vec{p})=-\frac {2\alpha_s}{3\pi^2}
\frac {1}{\vec{p}^{\;2}+\alpha ^2}.
\end{align}
Here, $\alpha_{s}$ is the strong coupling constant, the constants $\alpha$, $\lambda$ and $V_0$ are the parameters that characterize the potential.
In the following, we will employ this potential to both the $B_c$ system and the heavy-light quark system as an assumption.

In the numerical calculation, following parameters are used:
\begin{align}
&m_b=4.977\ {\rm GeV},\quad\quad m_c=1.628\ {\rm GeV},\quad\quad \alpha_s=0.21, \nonumber \\
&\alpha=0.06\ {\rm GeV},\quad\quad \lambda=0.315\ {\rm GeV}^2,\quad\quad V_0=-0.829\ {\rm GeV}. \nonumber
\end{align}
Here the heavy quark masses $m_b$ and $m_c$ are taken from Ref. \cite{Godfrey:2004ya}, the strong coupling constant $\alpha_s$ is taken from Ref. \cite{Godfrey:1985xj}.
The parameters $\alpha$, $\lambda$ and $V_0$ are fixed by fitting the mass spectrum to the latest experimental data \cite{Sirunyan:2019osb,Aaij:2019ldo}:
\begin{align}
&M(B_c(1^1S_0))=6271\ {\rm MeV},\quad \quad M(B_c(2^1S_0))=6872\ {\rm MeV}, \nonumber \\
&M(B_c^{*}(2^3S_1))-M(B_c^{*}(1^3S_1))=570\ {\rm MeV}.\nonumber
\end{align}
Based on the formalism and parameters above, we calculate the masses of $B_c(nS)$ states up to $n=4$.
The numerical results are shown in Table \ref{tab1}.
For comparison, results obtained from other approaches are also listed.
Note, since we fit to the latest experimental data of $B_c(2S)$ states, the masses of excited $B_c$ states of this work are generally larger than those of others.

\begin{table}
\begin{center}
\caption{Masses (MeV) of $B_c(nS)$ mesons.}
\label{tab1}
\begin{tabular}{|c|c|c|c|c|c|c|}\hline\hline
 State                            &     This work    & ~EQ~\cite{Eichten:1994gt}~~&~GI~\cite{Godfrey:2004ya}~~&{\scriptsize LLLLGZ}~\cite{Li:2019tbn}&AAMS~\cite{Asghar:2019qjl} & Lattice~\cite{Mathur:2018epb}\\ \hline						
$B_{c}(1 ^1S_{0})$     &6271       &            6264                     &            6271                    &       6271                          & 6318                                     &              6276              \\\hline
$B_{c}^*(1 ^3S_{1})$ &6346       &            6337                     &            6338                    &        6326                          & 6336                                       &              6331               \\\hline
$B_{c}(2 ^1S_{0})$     &6873                 &            6856                     &           6855                     &        6871                          & 6741                                   &               $\cdots$                      \\\hline
$B_{c}^*(2 ^3S_{1})$  &6916                &            6899                     &            6887                    &         6890                         & 6747                                                 &               $\cdots$                       \\\hline
$B_{c}(3 ^1S_{0})$      &7273                &            7244                     &           7250                     &         7239                         & 7014                                 &               $\cdots$                      \\\hline
$B_{c}^*(3 ^3S_{1})$  &7304                &            7280                     &            7272                     &         7252                         &7018                             &               $\cdots$                      \\\hline
$B_{c}(4 ^1S_{0})$      &7584                &            7562                     &             $\cdots$               &         7540                          &7239                             &                $\cdots$                     \\\hline
$B_{c}^*(4 ^3S_{1})$  &7606                &             7594                    &            $\cdots$                &          7550                         & 7242                             &               $\cdots$                    \\\hline
 \hline
\end{tabular}
\end{center}
\end{table}

\begin{figure}
\includegraphics[width=6.8cm]{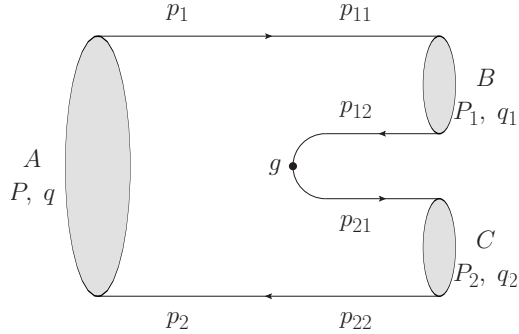}
\caption{The Feynman diagram of OZI-allowed two-body decay process with a $^3P_0$ vertex.} \label{ozi}
\end{figure}

Our results indicate that the $B_c(3S)$ states lie above the threshold for decay into a $BD$ meson pair.
The corresponding OZI-allowed two body decay can be depicted by $^3P_0$ model, where the additional light quark-antiquark pair is assumed to be created from vacuum, as shown in Fig. \ref{ozi}.
The usual $^3P_0$ model is a non-relativistic model with a transition operator $\sqrt{3}g\int d^3x\bar{\psi}(\vec{x})\psi(\vec{x})$, and it can be extended to a relativistic form $i\sqrt{3}g\int d^4 x\bar\psi(x)\psi(x)$ \cite{Simonov:2008cr,Wang:2013lpa}.
The coupling constant $g$ can be parameterized as $2m_q \gamma$, where $m_q$ is the constitute quark mass and $\gamma$ is a dimensionless parameter which can be extracted from experimental data.
Here we take $m_u=0.305\ {\rm GeV}$, $m_d=0.311\ {\rm GeV}$, and $\gamma=0.253\pm0.010$ \cite{Segovia:2012}.

The transition amplitude for the OZI-allowed two body decay process (with the momenta assigned as in Fig. \ref{ozi}) can be written as
\begin{equation}
\begin{aligned}
&\langle P_1P_2|S|P\rangle=(2\pi)^4\delta^4(P-P_1-P_2)\mathcal M \\
&= -ig\int\frac{d^4q}{(2\pi)^4}\frac{d^4q_1}{(2\pi)^4}\frac{d^4q_2}{(2\pi)^{4}}{\rm Tr}[\chi_{_P}(q)S^{-1}_2(p_2)(2\pi)^4\delta^4(p_2-p_{22})\bar\chi_{_{P_2}}(q_2)(2\pi)^4\delta^4(p_{12}-p_{21})\\
&~~~~\times \bar\chi_{_{P_1}}(q_1)S^{-1}_1(p_1)(2\pi)^4\delta^4(p_1-p_{11})]\\
&= -ig(2\pi)^4\delta^4(P-P_1-P_2)\int\frac{d^4q}{(2\pi)^4}{\rm Tr}[\chi_{_P}(q)S_2^{-1}(-p_2)\bar\chi_{_{P_2}}(q_2)\bar\chi_{_{P_1}}(q_1)S_1^{-1}(p_1)],
\end{aligned}
\end{equation}
where $q_i = q + (-1)^{i+1}\big(\tfrac{m_i}{m_1+m_2}P-\tfrac{m_{ii}}{m_{i1}+m_{i2}}P_i\big)$.
The Feynman amplitude takes the form~\cite{Wang:2013lpa}:
\begin{equation}\label{eq_FM}
\begin{aligned}
\mathcal M&=-ig\int\frac{d^4q}{(2\pi)^4}{\rm Tr}[\chi_{_P}(q)S_2^{-1}(-p_2)\bar\chi_{_{P_2}}(q_2)\bar\chi_{_{P_1}}(q_1)S_1^{-1}(p_1)]\\
&\approx g\int\frac{d^3 q_{\perp}}{(2\pi)^3} {\rm Tr} [\frac{\slashed P}{M}\varphi^{++}_{_P}(q_\perp)\frac{\slashed P}{M}\overline\varphi^{++}_{_{P_2}}(q_{2\perp})\overline\varphi^{++}_{_{P_1}}(q_{1\perp})].
\end{aligned}
\end{equation}
Note, to get the last line of Eq. \eqref{eq_FM}, we have used the fact that the wave function is strongly suppressed when $|q_\perp|$ is large.
With the method developed in Refs. \cite{Chang:2006tc,Zhang:2010iea}, the positive energy wave function can be determined by numerically solving Eq. \eqref{varphi++}.

The strong decay widths of $B_c(3S)$ mesons are shown in Table \ref{tab2}.
We obtain  $\Gamma\left(B_c(3^1S_0)\to B^*D\right)=26.02^{+2.33}_{-2.21}\ {\rm MeV}$, $\Gamma\left(B_c^*(3^3S_1)\to BD\right)=3.39^{+0.27}_{-0.26}\ {\rm MeV}$, $\Gamma\left(B_c^*(3^3S_1)\to B^*D\right)=14.77^{+1.40}_{-1.33}\ {\rm MeV}$ and $\Gamma\left(B_c^*(3^3S_1)\to BD^*\right)=6.14^{+0.58}_{-0.54}\ {\rm MeV}$.
In Table \ref{tab3}, we compare the strong decay widths with those of the non-relativistic $^3P_0$ model \cite{Ferretti:2015rsa,Li:2019tbn} and those of the Cornell coupled-channel model \cite{Eichten:2019gig}.
Our results are generally smaller than others, and we have following comments on the discrepancy:
\begin{enumerate}
\item In the $^3P_0$ model with  $\mathcal{H}_{\rm int}=i\sqrt{3}g\bar{\psi}(x)\psi(x)$, the transition operator  contains a factor $\frac{g}{2\omega_q}$, where $\omega_q$ is the energy of the created light quark.
This factor reduces to $\frac{g}{2m_q}$ in the non-relativistic limit.
Since the parameter $g$ is extracted from the fit of the non-relativistic $^3P_0$ model to the experimental data,
 our widths are in fact suppressed by $\frac{m_q^2}{\omega_q^2}$.
 By multiplying the compensation factor $\frac{\omega_q^2}{m_q^2}$, we find the decay widths are enhanced by a factor of about 3.
 In this sense, our results are compatible with those of Ref. \cite{Eichten:2019gig}.
\item In Ref. \cite{Ferretti:2015rsa} and Ref. \cite{Li:2019tbn}, the coupling constant $g$ is set to be about $0.292$ GeV and $0.264\ {\rm GeV}$, respectively, while here we tend to take the value of $0.155\ {\rm GeV}$ as given and discussed in Ref. \cite{Segovia:2012}.
This may lead to $3.5$ and 3 times difference in decay width, respectively.
 \item The relativistic effect of the wave functions is non-negligible, as discussed in Ref. \cite{Geng:2018qrl}.
 \item The masses of $B_c(3S)$ states are different from each other between our work, Ref. \cite{Ferretti:2015rsa} and Ref. \cite{Li:2019tbn}. This may lead to a deviation of about $20\%- 50\%$ in decay widths \cite{Eichten:2019gig}.
\end{enumerate}

\begin{table}
\begin{center}
\caption{Partial widths ($\Gamma$) and branching ratios (Br) of $B_c(3S)$ mesons. The number of events (NE) are estimated under the LHC experimental condition.}
\label{tab2}
\begin{tabular}{|c|c|c|c|c|c|c}\hline\hline
 Meson                           &   Decay mode                  & $\Gamma\ ({\rm {MeV}})$                           &   Br (\%)       & NE $(10^8)$\\ \hline
$B_c^{\pm}(3 ^1S_0)$ &  $B^\ast D^{\pm}$          &  $13.05^{+1.17}_{-1.11}$      &   $50.17$            & 2.45                    \\ \hline
$B_c^{\pm}(3 ^1S_0)$ &  $B^{\ast\pm} D^0$        &  $12.97^{+1.16}_{-1.10}$      &   $49.83$              &2.43                     \\ \hline
$B_c^{\pm}(3 ^3S_1)$ &  $B^0 D^{\pm}$             &  $1.76^{+0.11}_{-0.11}$      &   $7.25$              &1.04                  \\ \hline
$B_c^{\pm}(3 ^3S_1)$ &  $B^{\pm} D^0$             &  $1.63^{+0.16}_{-0.15} $     &   $6.71$              &0.969                      \\ \hline
$B_c^{\pm}(3 ^3S_1)$ &  $B^{\ast} D^{\pm}$     &   $7.24^{+0.72}_{-0.68}$      &   $29.79$                &4.30                   \\ \hline
$B_c^{\pm}(3 ^3S_1)$ &  $B^{\ast\pm} D^0$        &  $7.53^{+0.68}_{-0.65} $     &    $30.99$               &4.47                  \\ \hline
$B_c^{\pm}(3 ^3S_1)$ &  $B^{0} D^{\ast\pm}$     &   $2.73^{+0.26}_{-0.24}$      &   $11.21$                &1.62                   \\ \hline
$B_c^{\pm}(3 ^3S_1)$ &  $B^{\pm} D^{\ast0}$     &   $3.41^{+0.32}_{-0.30}$      &   $14.04$                &2.02                   \\ \hline
 \hline
\end{tabular}
\end{center}
\end{table}

\begin{table}
\begin{center}
\caption{Comparison of the results for $B_c(3S)$ strong decay widths\ ({\rm {MeV}}).
In Ref. \cite{Eichten:2019gig}, the decay widths of $B_c(3S)$ states as functions of their masses are presented.
 The results cited here are obtained by setting the $B_c(3S)$ masses to our predicted values.
 The $B_c(3S)$ masses in Ref. \cite{Ferretti:2015rsa} are $M(B_c(3^1S_0))=7249$ MeV and $M(B_c(3^3S_1))=7272$ MeV.
 The $B_c(3S)$ masses in Ref. \cite{Li:2019tbn} are shown in Table \ref{tab1}.}
\label{tab3}
\begin{tabular}{|c|c|c|c|c|c|c}\hline\hline
 Meson                           &   Decay mode                 & This work  &~EQ \cite{Eichten:2019gig}~      &~FS  \cite{Ferretti:2015rsa}~       &{\scriptsize LLLLGZ} \cite{Li:2019tbn}     \\ \hline
$B_c^{\pm}(3 ^1S_0)$ &  $B^\ast D$                     &  $26.02$           &65.92                                       &107                          &        161                                 \\ \hline
$B_c^{\pm}(3 ^3S_1)$ &  $BD$                            &  $  3.39$                &7.78                                      &13                            &          28                                \\ \hline
$B_c^{\pm}(3 ^3S_1)$ &  $B^{\ast} D$                &   $14.77$                  &35.18                                  &64                             &          105                              \\ \hline
$B_c^{\pm}(3 ^3S_1)$ &  $B D^{\ast}$                &   $6.14$                       &15.56                             &-                               &          -                                     \\ \hline
 \hline
\end{tabular}
\end{center}
\end{table}

According to non-relativistic quantum chromodynamics factorization formalism \cite{nrqcd}, the production rates of $B_c(3S)$ mesons can be estimated through
\begin{equation}
\sigma\left({B_c(3S)}\right)=\sigma\left({B_c(1S)}\right)\frac{|\Psi^{}_{B_c(3S)}(0)|^2}{|\Psi^{}_{B_c(1S)}(0)|^2},
\end{equation}
where $\Psi^{}_H(0)$ is the wave function at the origin for meson $H$.
With the $\sigma\left({B_c(1S)}\right)$ predicted in Ref. \cite{Chang:1992jb}, and the wave functions calculated in Ref. \cite{Eichten:1994gt}, the cross sections of $B_c(3S)$ mesons at the LHC can be estimated as $\sigma\left({B_c(3^1S_0)}\right)=4.88\ {\rm nb}$ and $\sigma\left({B_c^*(3^3S_1)}\right)=14.5\ {\rm nb}$.
At an integrated luminosity of $100\ \text{fb}^{-1}$, the numbers of $B_c(3^1S_0)$ and $B_c^\ast(3^3S_1)$ events are $4.88\times 10^8$ and $1.44\times10^9$, respectively.
The number of events for different decay channels are also presented in Table \ref{tab2}.

In summary, since the relativistic effects are evidently important in $B_c(3S)$ exclusive decays to $B$ and $D$ mesons, we have investigated in this work the mass spectrum and strong decay properties of $B_c(3S)$ states in the framework of BS equation and relativistic $^3P_0$ model. The numerical estimation gives $M(B_c(3^1S_0))=7273\ {\rm MeV}$, $M(B_c^*(3^3S_1))=7304\ {\rm MeV}$, $\Gamma\left(B_c(3^1S_0)\to B^*D\right)=26.02^{+2.33}_{-2.21}\ {\rm MeV}$, $\Gamma\left(B_c^*(3^3S_1)\to BD\right)=3.39^{+0.27}_{-0.26}\ {\rm MeV}$, $\Gamma\left(B_c^*(3^3S_1)\to B^*D\right)=14.77^{+1.40}_{-1.33}\ {\rm MeV}$ and $\Gamma\left(B_c^*(3^3S_1)\to BD^*\right)=6.14^{+0.58}_{-0.54}\ {\rm MeV}$.
We also estimate the number of events for different decay channels in the LHC experimental condition.
Since a large number of events may be produced in experiment, we suggest to find the $B_c(3S)$ states in their exclusive strong decays.

\newpage 
\vspace{0.0cm} {\bf Acknowledgments}

This work was supported in part by the National Natural Science Foundation of China (NSFC) under the Grants 11975236, 11635009, 12047553, 12075073, and by National Key Research and Development Program of China under Contracts Nos.2020YFA0406400.

\end{document}